\documentclass[aps,prl,reprint,superscriptaddress,notitlepage,a4]{revtex4-1}
\usepackage{graphicx}
\usepackage[rightcaption]{sidecap}
\usepackage{braket}
\usepackage[utf8]{inputenc}
\usepackage{setspace}
\usepackage{amsmath} 
\usepackage{amssymb}  
\usepackage{amstext}  
\usepackage{physics}
\graphicspath{{./figures/}}

\begin{document}
\title{Optical spin locking of a solid-state qubit}

\author{J.H. Bodey}
\affiliation{Cavendish Laboratory, University of Cambridge, JJ Thomson Avenue, Cambridge, CB3 0HE, UK}

\author{R. Stockill}
\email[Present address: ]{Kavli Institute of Nanoscience, Delft University of Technology, Lorentzweg 1, 2628 CJ Delft, The Netherlands}
\affiliation{Cavendish Laboratory, University of Cambridge, JJ Thomson Avenue, Cambridge, CB3 0HE, UK}

\author{E.V. Denning}
\affiliation{Cavendish Laboratory, University of Cambridge, JJ Thomson Avenue, Cambridge, CB3 0HE, UK}
\affiliation{Department of Photonics Engineering, Technical University of Denmark, 2800 Kgs. Lyngby, Denmark}

\author{D.A. Gangloff}
\affiliation{Cavendish Laboratory, University of Cambridge, JJ Thomson Avenue, Cambridge, CB3 0HE, UK}

\author{G. \'Ethier-Majcher}
\affiliation{Cavendish Laboratory, University of Cambridge, JJ Thomson Avenue, Cambridge, CB3 0HE, UK}

\author{D.M. Jackson}
\affiliation{Cavendish Laboratory, University of Cambridge, JJ Thomson Avenue, Cambridge, CB3 0HE, UK}

\author{E. Clarke}
\affiliation{EPSRC National Epitaxy Facility, University of Sheffield, Sheffield, Broad Lane, S3 7HQ, UK}

\author{M. Hugues}
\affiliation{Universit\'e C\^ote d'Azur, CNRS, CRHEA, Valbonne, France}

\author{C. Le Gall}
\email[Electronic address: ]{cl538@cam.ac.uk}
\affiliation{Cavendish Laboratory, University of Cambridge, JJ Thomson Avenue, Cambridge, CB3 0HE, UK}

\author{M. Atat\"ure}
\email[Electronic address: ]{ma424@cam.ac.uk}
\affiliation{Cavendish Laboratory, University of Cambridge, JJ Thomson Avenue, Cambridge, CB3 0HE, UK}

\begin{abstract}
Quantum control of solid-state spin qubits typically involves pulses in the microwave domain, drawing from the well-developed toolbox of magnetic resonance spectroscopy. Driving a solid-state spin by optical means offers a high-speed alternative, which in the presence of limited spin coherence makes it the preferred approach for high-fidelity quantum control. Bringing the full versatility of magnetic spin resonance to the optical domain requires full phase and amplitude control of the optical fields.  Here, we imprint a programmable microwave sequence onto a laser field and perform electron spin resonance in a semiconductor quantum dot via a two-photon Raman process. We show that this approach yields full SU(2) spin control with over $98\%$  $\pi$-rotation fidelity.  We then demonstrate its versatility by implementing a particular multi-axis control sequence, known as spin locking. Combined with electron-nuclear Hartmann-Hahn resonances which we also report in this work, this sequence will enable efficient coherent transfer of a quantum state from the electron spin to the mesoscopic nuclear ensemble.
\end{abstract}
\pacs{}
\date{\today}
\maketitle
\section{Introduction}
The existence of strong electric dipole transitions enables coherent optical control of matter qubits that is both fast and local \cite{Arimondo1976,Meekhof1996,Johnson2008}. The optical techniques developed to address central spin systems in solids, such as colour centers in diamond and confined spins in semiconductors, typically fall into two categories: the first makes use of ultrashort, broadband, far-detuned pulses to induce quasi-instantaneous qubit rotations in the laboratory frame \cite{Gupta2001, Press2008, Campbell2010, Becker2016}. Achieving complete quantum control with this technique further requires precisely timed free qubit precession accompanying the optical pulses. The second technique is based on spectrally selective control via a resonant drive of a two-photon Raman process \cite{Golter2014,Delley2017,Zhou2017,Becker2018,Goldman2018}, and allows full control exclusively through tailoring of the drive field, echoing the versatility of magnetic spin resonance. Despite this attractive flexibility, achieving high-fidelity control using the latter approach has proved challenging due to decoherence induced by the involvement of an excited state for colour-centers in diamond \cite{Golter2014,Zhou2017,Becker2018}, and due to nuclei-induced ground-state decoherence for optically active semiconductor quantum dots (QDs) \cite{Delley2017}. In the case of QDs, the limitation of ground-state coherence can be suppressed by preparing the nuclei in a reduced-fluctuation state \cite{Xu2009,Bluhm2010,Issler2010,Ethier-Majcher2017,Gangloff2019}. In this Letter, we achieve high-fidelity SU(2) control on a nuclei-prepared QD spin using a tailored waveform imprinted onto an optical field.  We then demonstrate the protection of a known quantum-state via an aligned-axis continuous drive, a technique known as spin locking. Finally, by tuning the effective spin Rabi frequency, we access the electron-nuclear Hartmann-Hahn resonances which holds promise for proxy control of nuclear states.\\
\section{Results}
\subsection{Optical electron spin resonance}
\begin{figure}
\includegraphics[width = 1\columnwidth,angle=0]{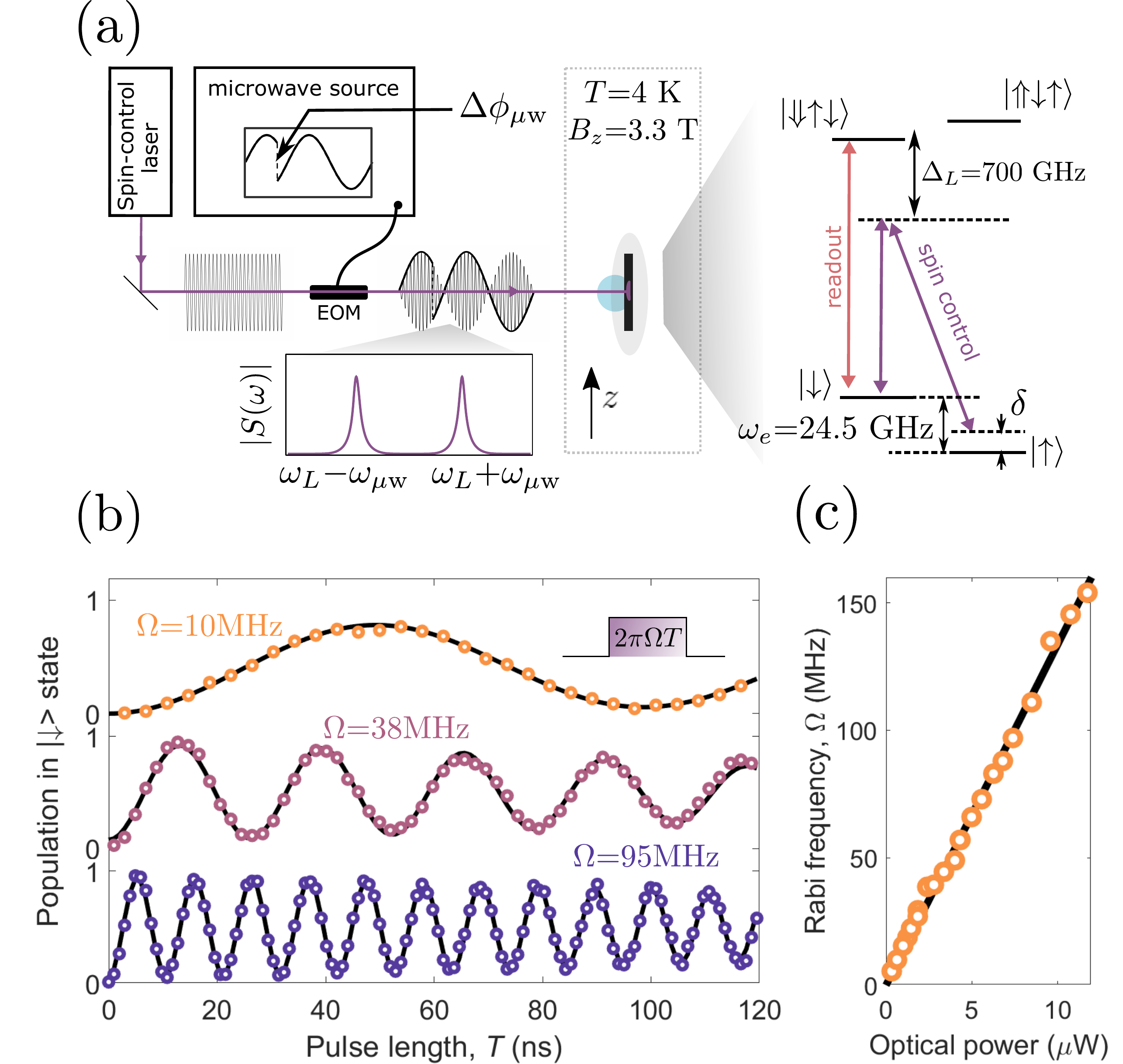}
\caption{All-optical electron spin resonance (ESR). (a) Experimental schematic: Intensity modulation of a single-frequency laser produces two sidebands for spin-control. Encoding a phase step $\Delta \phi_{\mathrm{\mu w}}$ in the microwave signal produces a change of relative phase $\phi = 2 \Delta \phi_{\mathrm{\mu w}}$ between these two sidebands. These then drive two-photon Raman transitions between the energy levels of a negatively charged QD, as shown on the right. The optical fields have a single-photon detuning from the excited state of $\Delta_L\approx700\mathrm{\ GHz}$, and a two-photon detuning from the ESR of $\delta$. A resonant laser pulse is used to initialise the spin via optical pumping prior to spin control and to read out the population of the $\ket{\downarrow}$ state after spin control. (b) Time evolution of $\ket{\downarrow}$-state population after drive-time T taken at three different Rabi frequencies. The solid curves are fits from a Bloch-equation model to extract the Rabi frequency $\Omega$. (c) Dependence of the  Rabi frequency $\Omega$ on the laser power incident on the cryostat window. The black line is a linear fit, with a slope of $13.4\mathrm{\ MHz.} \mu \mathrm{W}^{-1}$.}
\label{fig:1}
\end{figure}
\indent Our device is an Indium Gallium Arsenide QD, embedded in an n-type Schottky heterostructure and housed in a liquid-helium cryostat at $4.2\ \mathrm{K}$; Figure \ref{fig:1} (a) depicts this arrangement. The QD is charged deterministically with a single electron, and a magnetic field of $3.3\ \mathrm{T}$ perpendicular to the growth and optical axes creates a  $\omega_e=24.5\ \mathrm{GHz}$ Zeeman splitting of the electron spin states which form $\Lambda$ systems with the two excited trion states. Using an electro-optic modulator (EOM), we access these $\Lambda$ systems by tailoring a circularly polarised single-frequency laser, of frequency $\omega_L$ and detuned from the excited states by $\Delta_L\approx700\ \mathrm{GHz}$. The EOM is driven by an arbitrary waveform generator output with amplitude $V_{0}$, frequency $\omega_{\mathrm{\mu w}}$ and phase $\phi_{\mathrm{\mu \mathrm{w}}}$. Operating the EOM in the regime where the microwave field linearly modulates the input optical field,  a signal $V_0 \cos(\omega_{\mathrm{\mu \mathrm{w}}} t+\Delta \phi_{\mathrm{\mu \mathrm{w}}})$ produces a control field consisting of two frequencies at $\omega_{L}\pm\omega_{\mathrm{\mu w}}$ with a relative phase-offset of $2 \Delta \phi_{\mathrm{\mu w}}$. This bichromatic field of amplitude $\Omega_L$ drives the two-photon Raman transitions with a Rabi coupling strength $\Omega=\Omega_L^2/\Delta_L$ between the electron spin states \cite{Supplementary} in the limit $(\Omega_L/\Delta_L)^2\ll1$. The Hamiltonian evolution is given by:
\begin{align*}
\hat{\mathcal{H}}_{\mathrm{eff}}=\frac{\Omega}{2}(\cos(\phi)\hat{\sigma}_x+\sin(\phi)\hat{\sigma}_y)+ \frac{\delta}{2} \hat{\sigma}_z
\end{align*}  
\noindent where $\hat{\sigma}_i$ are the Pauli operators in the electron rotating frame, $\delta$ the two-photon detuning and $\phi$ the relative phase-offset of the Raman beams. The effect of this Hamiltonian is described geometrically by a precession of the Bloch vector around the Rabi vector [$\Omega \cos(\phi), \Omega \sin(\phi), \delta$]. We have full SU(2) control over the Rabi vector through the microwave waveform, via the Rabi frequency $\Omega\propto V_0^2$, its  phase $\phi=2 \Delta \phi_{\mu \mathrm{w}}$, and the two-photon detuning $\delta=\omega_{e}-2\omega_{\mu \mathrm{w}}$. An additional resonant optical field of $100$-$\mathrm{ns}$ duration performs spin initialisation and read-out. Finally, prior to the whole protocol, we implement the recently developed nuclear-spin narrowing scheme \cite{Gangloff2019}, which conveniently requires no additional laser or microwave source, in order to enhance ground-state coherence and so maximise control fidelity.\\
\indent Figure \ref{fig:1} (b) shows the evolution of the population of the $\ket{\downarrow}$ state for increasing durations of the Raman drive, taken at three different Raman powers. The Raman drive induces coherent Rabi oscillations within the ground-state manifold. The dependence of the fitted Rabi frequency on power is linear within the power range experimentally available as shown in Fig. \ref{fig:1} (c). This linearity is the result of modest optical power ($\sim 10\mathrm{\ \mu W}$) and a sufficiently large single photon detuning $\Delta_L\approx700\mathrm{\ GHz}$, allowing us to work in the adiabatic limit where excited-state population is negligible during the rotations. Even in this limit, we reach Rabi frequencies up to $154\mathrm{\ MHz}$, exceeding that achieved by extrinsic spin-electric coupling \cite{Yoneda2018, Zajac2018} and two orders of magnitude faster than direct magnetic control of gate-defined spin qubits \cite{Veldhorst2014}.  While rotations driven by ultrafast (few ps), modelocked-laser pulses naturally circumvent ground-state dephasing, the high visibility of the Rabi oscillations achieved here suggest that our electron spin resonance (ESR) yields equally coherent rotations with the added spectral selectivity and flexibility of microwave control.\\
\begin{figure}
\includegraphics[width = 0.95\columnwidth,angle=0]{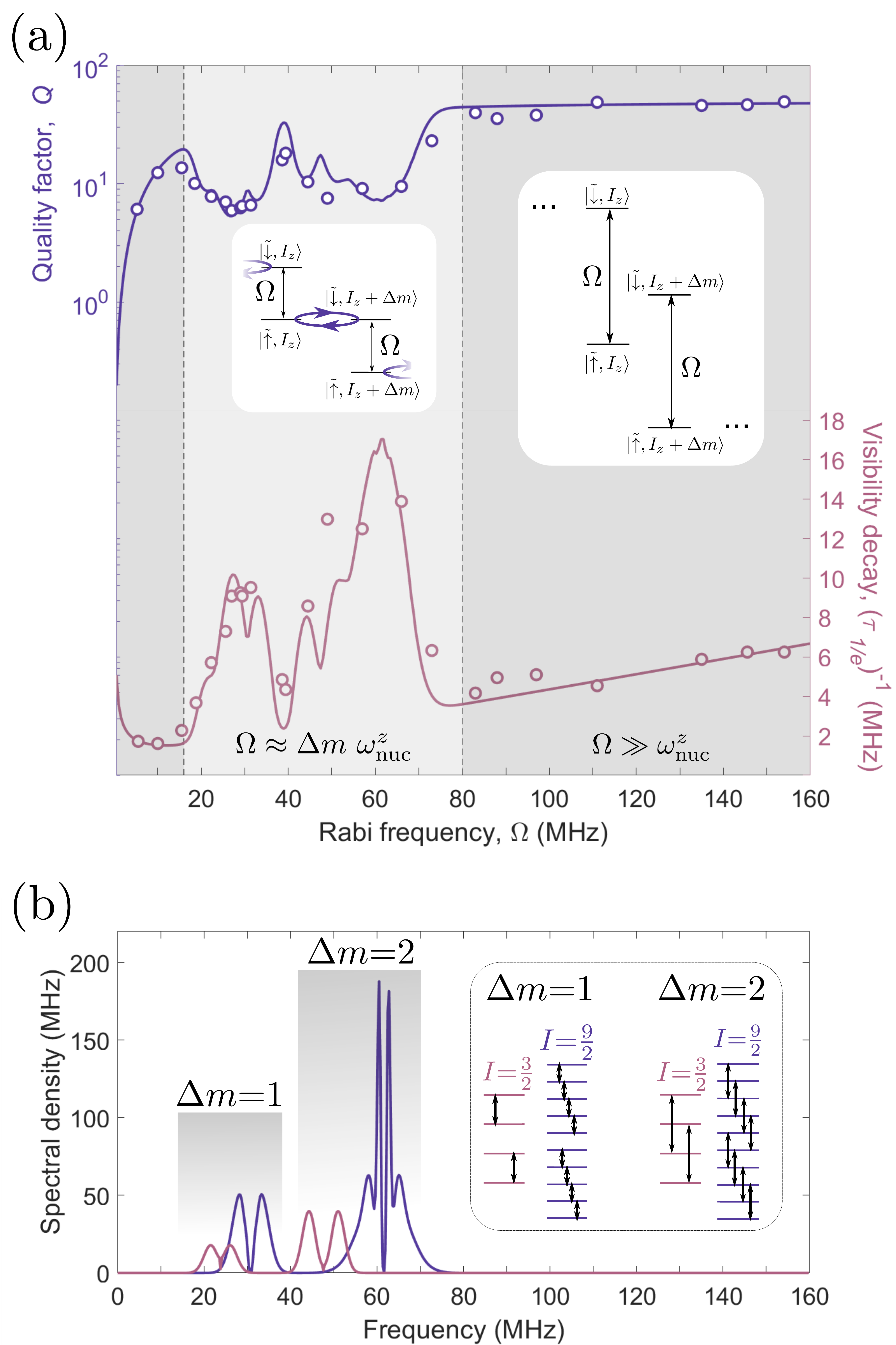}
\caption{ESR properties of the driven central spin. (a) $Q$-factor of the Rabi oscillations (purple) and inverse $1/e$-decay time of the Rabi oscillation visibility (pink) as a function of Rabi frequency. We define the visibility by taking the maximum and minimum of the Rabi curve, over a $\pi$-period. Hartmann-Hahn resonances between electron-nuclear states at $\Omega \approx \omega^z_{\mathrm{nuc}}$ and $\Omega\approx 2 \omega^z_{\mathrm{nuc}}$ depicted in the inset emerge as an accelerated decay. (b) Calculated nuclear spectral density for Indium (blue, $I$=9/2) and Arsenic (pink, $I$=3/2). The inset indicates the transitions $\Delta m=1$ and $\Delta m=2$ strain-allowed to first order, considered in our model. In the intermediate-power regime where $\Omega \approx \Delta m\ \omega^z_{\mathrm{nuc}}$, $(\tau_{1/e})^{-1}$ is proportional to this density of states, after convolution with the width of the ESR transition \cite{Supplementary}.}
\label{fig:2} 
\end{figure}
\subsection{Coherence of optical rotations}
\indent We characterise the coherence of the rotations with the quality factor $Q$, which measures the number of $\pi$ rotations before the Rabi-oscillation visibility falls below $1/e$ of its initial value. Figure 2 (a) summarises the dependence of the $Q$ factor and decay of the Rabi envelope on the ESR drive strength $\Omega$ and sheds light on three distinct regimes which are dominated by one of three competing decoherence processes included in the model curve of Fig. \ref{fig:2}: (i) inhomogeneous broadening of variance $\sigma=4.8 \mathrm{\ MHz}$ (ii) electron-mediated nuclear spin-flipping transitions arising from the presence of strain (iii) a spin decay proportional to the laser power, which for simplicity we cast as  $\Gamma_1=\alpha|\Omega|$ with $\alpha=2.7\times10^{-2}$. In the low-power regime, where $\Omega <18\ \mathrm{MHz}$, the fidelity is affected by nuclei-induced shot-to-shot detuning errors. This inhomogeneous broadening induces a non-exponential decay of Rabi oscillation visibility \cite{Johnson2008,Supplementary}.  Increasing the Rabi frequency shields the system from this effect, yielding an increase in $Q$ factor. The intermediate-power regime, where $\Omega = 18-80\ \mathrm{MHz}$, exhibits a dramatic decrease in $Q$ and increased decay rate. In this regime, the coherent spectrally-selective drive induces electron-mediated nuclear spin-flips \cite{Gangloff2019} through a Hartmann-Hahn resonance \cite{Hartmann1962}, as we depict in the inset to Fig. \ref{fig:2}(a). Splitting the dressed electron states $\tilde{\ket{\uparrow}}, \tilde{\ket{\downarrow}}$ by an energy $\hbar \Omega$ causes the dressed electron-nuclear states to become degenerate, removing the energy cost associated to a single nuclear spin-flip $\sim\hbar \omega^z_{\mathrm{nuc}}$. The presence of intrinsic strain, which perturbs the nuclear quantisation axis set by the external magnetic field, allows coupling between these now-degenerate states  \cite{Supplementary}. The decay of electronic coherence is related to the nuclear spectral density shown in Fig. \ref{fig:2}(b), which captures the strength of the strain-enabled nuclear transitions. In the high-power regime ($\Omega>80\mathrm{\ MHz}$), we decouple from both inhomogeneous nuclear spin fluctuations and Hartmann-Hahn transitions, and consequently observe the highest $Q$ factors ($Q=47.6\pm1.7$ averaged over the four highest Rabi frequencies). Here, the decay envelope is dominated by $\Gamma_1$, an optically induced relaxation between the electron states proportional to power, and independent of detuning \cite{Supplementary}. The non-resonant and non-radiative nature of this process is consistent with electron-spin relaxation induced by photo-activated charges appearing  in our device as a DC Stark shift of the resonance \cite{Houel2012}. This mechanism, extrinsic to the QD, will vary depending on device structure \cite{Houel2012, Ding2018} and quality. This process causes an exponential decay of the Rabi oscillations, presenting a theoretical upper bound on the $Q$ factor of $4/(3\alpha) = 49$ and on the $\pi$-rotation fidelity $f_{\pi}=\frac{1}{2}\times (1+e^{-1/Q})$ of $0.989$ \cite{Supplementary}. Our model also allows us to evaluate the correction to this bound (of order $10^{-3}$) due to the non-Markovian effects of the nuclear inhomogeneities and Hartmann-Hahn resonances within the spectral width $1/t_\pi=2\Omega$ of the $\pi$ pulse. As a result, our highest $\pi$-pulse fidelity, measured at $\Omega=154 \mathrm{\ MHz}$, is $f_{\pi}=0.9886(4)$.\\
\subsection{Multi-axis control}
\begin{figure}
\includegraphics[width = 1\columnwidth,angle=0]{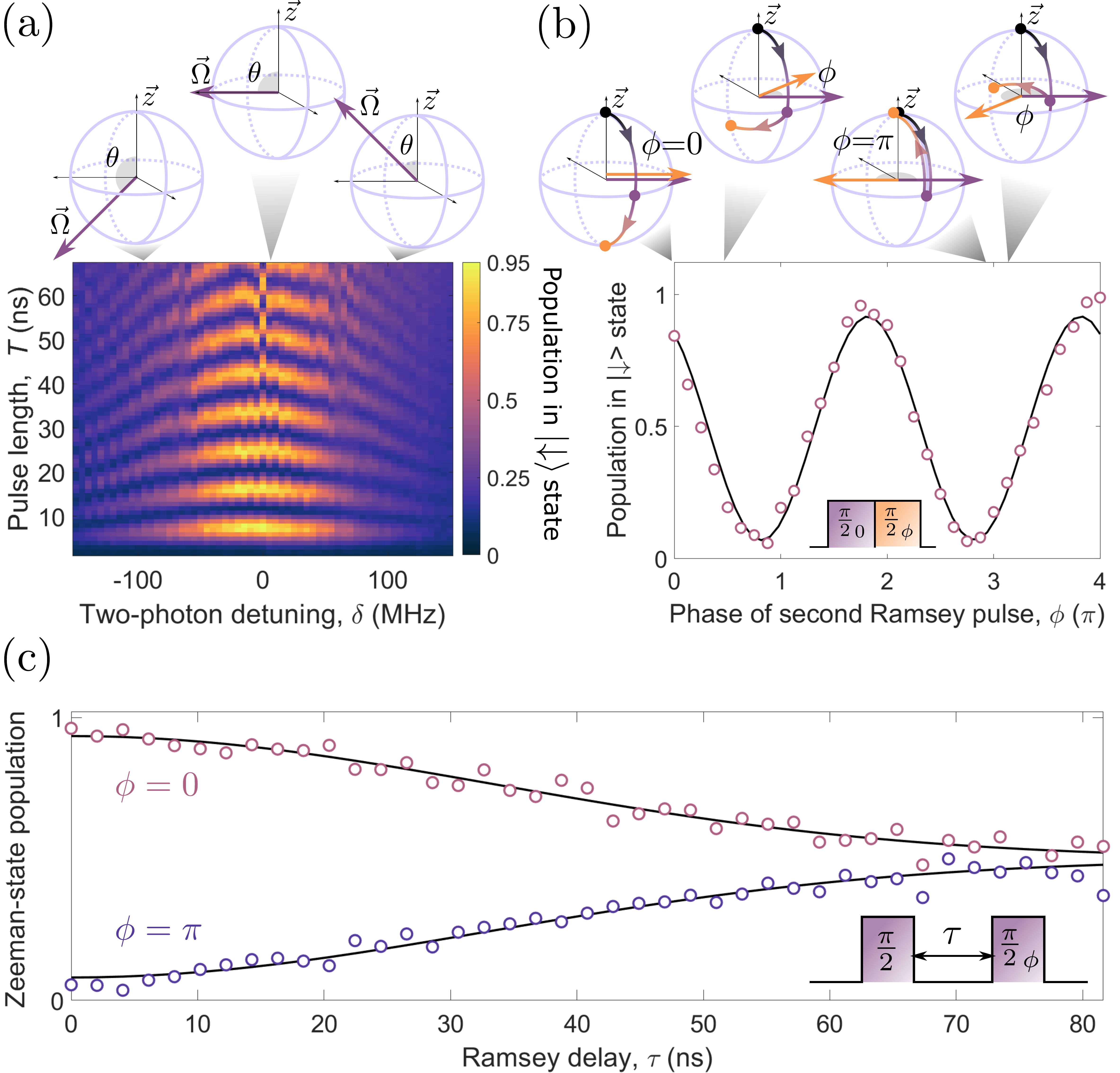}
\caption{SU(2) control over the rotation axis. (a) Rabi oscillations as a function of $\delta$, at a bare Rabi frequency $\Omega=120\ \mathrm{MHz}$. $\delta$ dictates the polar angle $\theta=\mathrm{arctan}(\frac{\Omega}{\delta})$ of the Rabi vector. (b) Dependence of the $\ket{\downarrow}$-state population on the relative phase $\phi$ of two immediately-consecutive $13\ \mathrm{MHz}\ \frac{\pi}{2}$-pulses, as $\phi_{\mu \mathrm{w}}$ is varied between $[0, 2\pi]$. This phase corresponds to the azimuthal angle of the Rabi vector. The phase offset between maximum readout signal and constructive pulse interference is consistent with a systematic detuning of $3.5\ \mathrm{MHz}$. (c) Ramsey interferometry on the electron. Two $24 \mathrm{MHz}\ \frac{\pi}{2}$-pulses, separated by a delay $\tau$ and with $\phi=0$ ($\phi=\pi$), produce the pink (purple) data points. These data are fitted by a Gaussian envelope, $\rho(t)= \frac{\rho_0}{2}(1\pm  e^{-(t/T_2^*)^2})$ for an initial population $\rho_0$, yielding a $47.4\ (47.1)$-$\mathrm{ns}$ inhomogeneous dephasing time for the upper (lower) curve. }
\label{fig:3}
\end{figure}
\indent Figure \ref{fig:3}(a) shows Rabi oscillations taken while varying the detuning $\delta$. With increasing detuning $\abs{\delta}$, the frequency of the Rabi oscillations $\Omega'=\sqrt{\Omega^{2} + \delta^{2}}$ increases, while the amplitude $\frac{\Omega^{2}}{\Omega'^{2}}$ decreases, as the spin precession follows smaller circles on the Bloch sphere. This confirms that we control the polar angle $\theta$ of the Rabi vector through detuning of the microwave field.\\
\indent In Fig. \ref{fig:3}(b), we demonstrate control over the azimuthal angle of the rotation axis by stepping the phase $\phi$ between two consecutive $\frac{\pi}{2}$ rotations. The $\ket{\downarrow}$-state population evolves sinusoidally with the phase shift between the two $\frac{\pi}{2}$ pulses. For example, at  $\phi=0$, the two rotations add resulting in a $\pi$ rotation and maximum readout signal, whilst for $\phi=\pi$, the two pulses exactly cancel, returning the electron spin to its starting state and giving a minimum readout signal. Defining the measurement as the $\frac{\pi}{2}_{\phi}$ pulse combined with the $\ket{\downarrow}$-state readout, the phase dependence shown here demonstrates our ability to perform $\sigma_{\pm x}$ and $\sigma_{\pm y}$ measurements, corresponding to two-axis tomography.\\ 
\indent Figure \ref{fig:3}(c) displays Ramsey interferometry performed in the rotating frame, which allows us to further characterise our ESR control. We create a spin superposition using a resonant $\frac{\pi}{2}$ pulse, which evolves for a time $\tau$ before measuring the state using a second $\frac{\pi}{2}$ pulse  with a relative phase $\phi=0$ ($\phi=\pi$), performing a $\sigma_y$ ($\sigma_{-y}$) measurement. Within this observation window, there are no oscillations modulating the dephasing-induced decay ($T_2^*$), confirming that the measurement basis is phase-locked to the rotating frame to below our resolution, set by the inhomogeneous nuclear broadening. Under these optimum nuclear spin narrowing conditions \cite{Supplementary, Gangloff2019}, the spin coherence decays according to $T_2^*=47.2\pm0.2\ \mathrm{ns}$; this corresponds to a standard deviation of the spin splitting of $\sigma=4.77\pm0.02\ \mathrm{MHz}$ due to the hyperfine fluctuations.\\
\begin{figure}
\includegraphics[width = 1\columnwidth,angle=0]{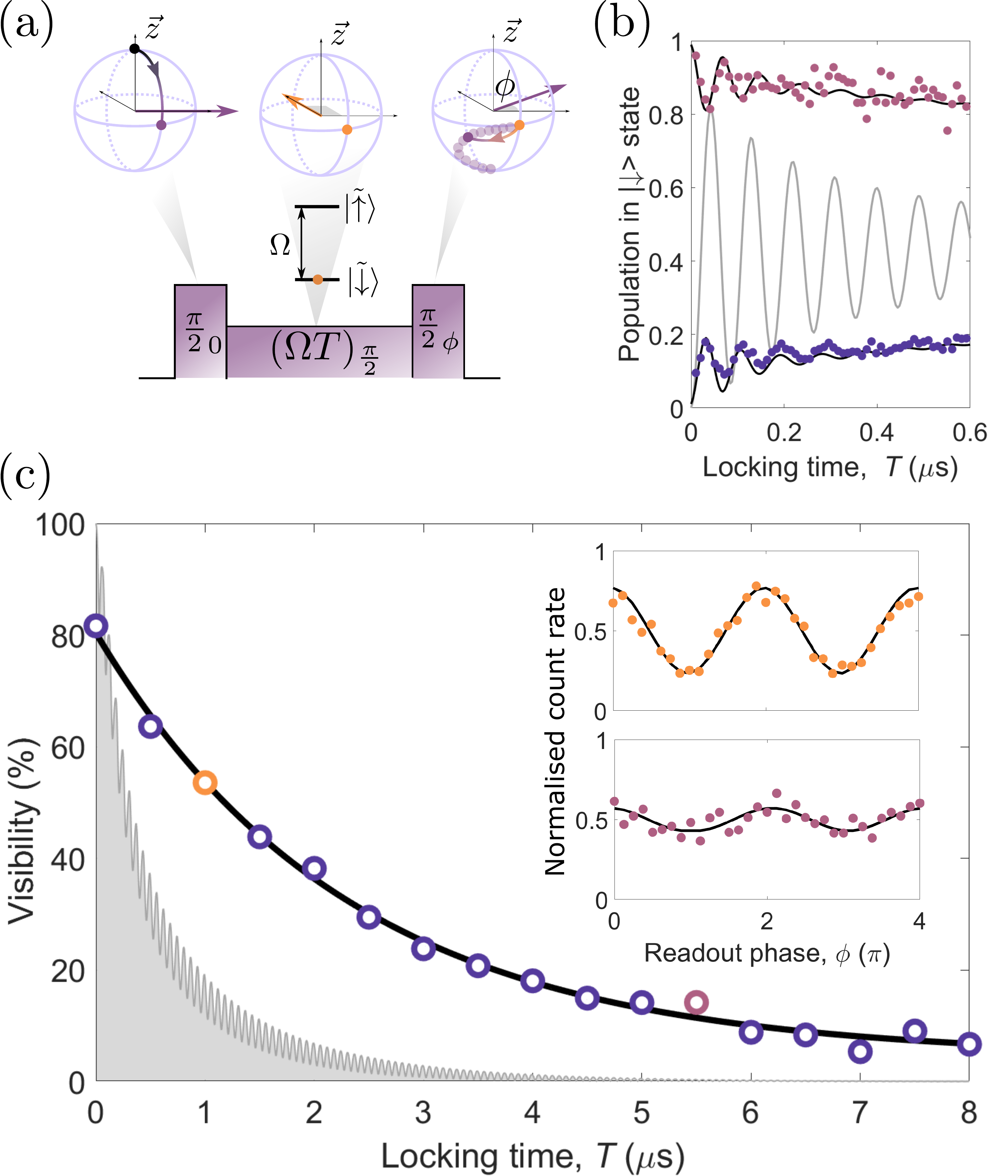}
\caption{Optical locking of a coherent superposition. (a) Spin-locking sequence schematic in the rotating frame. The electron, initially in $\ket{\uparrow}_z$ state (black dot), is rotated to $\tilde{\ket{\downarrow}}=\ket{\downarrow}_{y}$ (purple dot) by the first $\frac{\pi}{2}$ pulse. The phase of the drive is then jumped by $\frac{\pi}{2}$: $\tilde{\ket{\downarrow}}$ (yellow dot) is now an eigenstate of the drive, protected from dephasing by an energy gap $\Omega$. The system is driven in this configuration for a time $T$. A final $\frac{\pi}{2}$ pulse with phase $\phi$ before the $\ket{\downarrow}_{z}$ readout allows the equatorial spin components to be measured. (b) Spin locking with $\Omega=11\mathrm{\ MHz}$ as a function of locking time $T$, with a readout phase of $\phi=0\ (\pi)$ producing the pink (purple) data. The data are presented alongside a Bloch-equation model (black line: spin-locking, grey line: direct Rabi drive) that accounts for the inhomogeneous broadening of $\sigma=4.8\mathrm{\ MHz}$ and spin decay $\Gamma_1$. (c) Spin locking at  $\Omega=16\mathrm{\ MHz}$ as a function of locking time $T$. Tomography of the state in the $xy$ plane is done by varying the phase $\phi$ of the final $\frac{\pi}{2}$ pulse over $4\pi$ after each locking time; the insets depict two such datasets, indicated by colour. We use these data to extract a visibility, fitted with an exponential decay time of $2.3\pm 0.2\ \mu\mathrm{s}$ (black line). The corresponding visibility for a direct Rabi drive is plotted alongside (grey line) and exhibits decay on a $100$-$\mathrm{ns}$ timescale. Nuclear-field inhomogeneities lead to the oscillations seen in the Rabi visibility, which is partially refocussed at integer multiples of a $2 \pi$ rotation.}
\label{fig:4} 
\end{figure}
An immediate opportunity derived from multi-axis control is the realisation of an optical analogue of spin locking, an established magnetic resonance sequence designed to preserve a known quantum state well beyond its dephasing time.  In this sequence [Fig. \ref{fig:4}(a)], a $\frac{\pi}{2}$ rotation creates the quantum state $(\ket{\uparrow}-\mathrm{i}\ket{\downarrow})/\sqrt{2} $ in the equatorial plane, which has a dephasing time of $47.2\ \mathrm{ns}$. The azimuthal angle of the rotation axis is then shifted by $\frac{\pi}{2}$, bringing the Rabi vector into alignment with the system state; this places the electron into one of the dressed states. The drive creates an energy gap $\Omega$ between the two dressed states, which provides protection against environmental dynamics occurring at frequencies different from $\Omega$. By setting the gap size $\Omega$ to $\sim 10 \ \mathrm{MHz}$, we successfully avoid nuclear-spin resonances observed in Fig. \ref{fig:2}. Figure \ref{fig:4}(b) displays the population in the \{${\tilde{\ket{\uparrow}},\ \tilde{\ket{\downarrow}}}$\} basis during the first $600\ \mathrm{ns}$ of the spin-locking sequence. At these short delays, a small unlocked component of the Bloch vector undergoes Rabi oscillations resulting in small-amplitude oscillations. As confirmed with our Bloch-equation model [black curve in Fig. \ref{fig:4}(b)], this arises from detuning errors of the locking pulse consistent with the measured 4.8-MHz nuclear field inhomogeneity. The decay of the locked component of the Bloch vector is significantly slower than under a Rabi drive of the same amplitude ($\Omega=11\mathrm{\ MHz}$) [grey model curve in Fig. \ref{fig:4}(b)].
Figure \ref{fig:4}(c) shows the decay of the spin-locked state on longer timescales. After each locking window, at $\Omega=16\ \mathrm{MHz}$, we measure the length of the Bloch vector by performing state tomography and obtaining the visibility as in Fig. \ref{fig:3}(b).  An exponential fit  [black curve in Fig. \ref{fig:4}(b)] reveals a decay time of $2.3 \pm 0.2\ \mathrm{\mu s}$. The close agreement with the decay rate expected from our Fig. \ref{fig:2} model is evidence that spin locking is similarly limited by the photo-activated spin relaxation ($\Gamma_1$). The quantum state $(\ket{\uparrow}-\mathrm{i}\ket{\downarrow})/\sqrt{2}$ is thus preserved for a thousand times longer than the bare dephasing time, fifty times longer than the cooled-nuclei dephasing time, and three times longer than with direct Rabi drive. \\
\section{Discussion}
\indent The high-fidelity all-optical ESR we report here enables the generation of any quantum superposition spin state on the Bloch sphere using a single waveform-tailored optical pulse. This full SU(2) control further allows the all-optical implementation of spin locking, traditionally an NMR technique, for quantum-state preservation via gapped protection from decoherence-inducing environmental dynamics. In the case of semiconductor QDs, where the nuclei form the dominant noise source, the same quantum control capability enables us to reveal directly the spectrum of nuclear-spin dynamics. An immediate extension of this work will be to perform spin locking in the spectral window of nuclear-spin resonances, i.e. the Hartmann-Hahn regime, to sculpt collective nuclear-spin states \cite{Reynhardt1998,Henstra2008}, and also to tailor the electron-nuclear interaction \cite{Malinowski2016,Abobeih2018,Schwartz2018} to realise an ancilla qubit or a local quantum register based on the collective states of the nuclear ensemble \cite{Denning2019}.\\
\section{Methods}
\subsection{Quantum dot device}
Our QD device is the one used in Ref. \cite{Stockill2016}. Self-assembled InGaAs QDs are grown by Molecular Beam Epitaxy and integrated inside a Schottky diode structure, above a distributed Bragg reflector to maximize photon-outcoupling efficiency. There is a $35$-$\mathrm{nm}$ tunnel barrier between the n-doped layer and the QDs, and a tunnel barrier above the QD layer to prevent charge leakage. The Schottky diode structure is electrically contacted through Ohmic AuGeNi contacts to the n-doped layer and a semitransparent Ti gate ($6\ \mathrm{nm}$) is evaporated onto the surface of the sample. The photon collection is enhanced with a superhemispherical cubic zirconia solid immersion lens (SIL) on the top Schottky contact of the device. We estimate a photon-outcoupling efficiency of 10\% at the first lens for QDs with an emission wavelength around $970\ \mathrm{nm}$. A home-built microscope with spectral and polarisation filtering \cite{Supplementary} is used for resonance fluorescence, with a QD-to-laser counts ratio exceeding 100:1.
\subsection{Raman laser system}
\begin{figure}
\includegraphics[width = 1\columnwidth,angle=0]{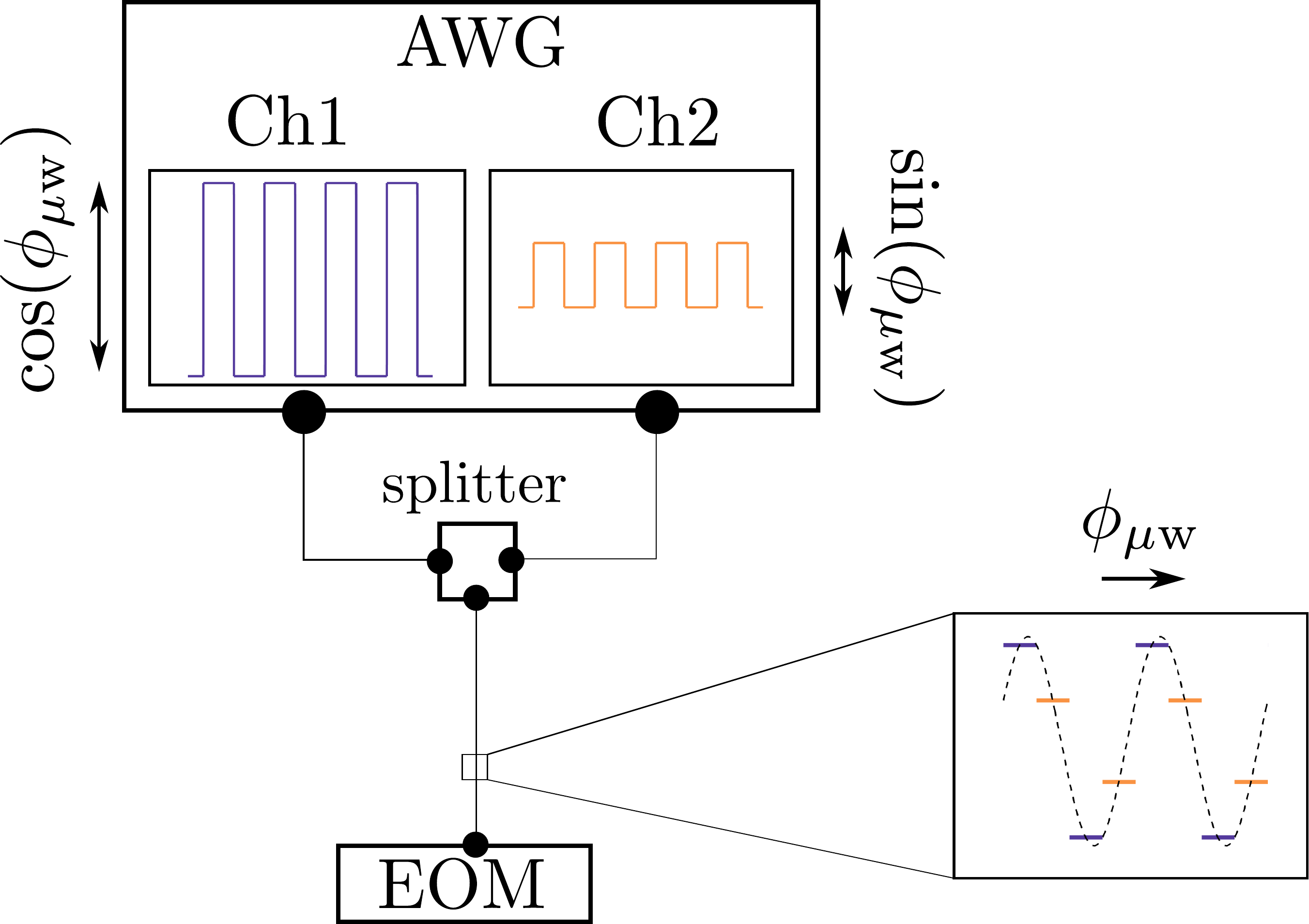}
\caption{Schematic of our microwave-generation apparatus. Two channels of an AWG produce square waves with an amplitude ratio of $\tan{\phi_{\mu \mathrm{w}}}$. These square waves are mixed in quadrature to construct a phase-controlled sine-wave, with a phase $\phi_{\mu \mathrm{w}}$.}
\label{fig:Splitter}
\end{figure}

Sidebands are generated from the continuous-wave (CW) laser by modulating a fibre-based EOSPACE electro-optic modulator (EOM) with a microwave derived from a Tektronix Arbitrary Waveform Generator (AWG) 70002A. The electric field at the EOM output $E$ is described by $E_{out}(t) \propto V_{in}(t) \times E_{in}(t)$ for an applied voltage $\abs{V_{in}}<<\abs{V_{\pi}}$. In other words, we work with small amplitude around the minimum intensity transmission of the EOM.\\
\indent Generation of the microwave signal $V_{in}(t)$ is depicted in Fig. \ref{fig:Splitter}. We produce a digital signal with a sampling rate that is four times the microwave frequency (a factor 2 is obtained by setting the AWG sampling rate at $2\omega_{\mu \mathrm{w}}$ and another factor 2 is obtained by combining two independently programmable AWG outputs with a splitter). We thus arrive at a digital signal containing four bits per period, the minimum required to carry phase information to the EOM. To generate the signal shown in Fig. \ref{fig:Splitter}, we add the two AWG outputs in quadrature, which we realize after characterisation of the relative delay between the two microwave lines arriving at the splitter. From each output, we generate a square-wave signal at $12.25\ \mathrm{GHz}$. By tuning their relative amplitudes, we construct a digitised sinusoidal signal at $12.25\ \mathrm{GHz}$ whose phase $\phi$ is determined by the relative amplitudes $A_{1,2}$ of channels 1 and 2 according to tan($\phi$)=$A_{1}/A_{2}$.
\subsection{Experimental cycle}
\subsubsection{Nuclear-spin preparation}
Figure \ref{fig:PulseSequence} shows our experimental cycle which involves narrowing the nuclear-spin distribution before a spin-manipulation experiment. Nuclear-spin preparation is done using the scheme detailed in Ref. \cite{Gangloff2019}, operating in a configuration analogous to Raman cooling in atomic systems. It involves driving the system continuously with the Raman laser, while pumping the $\ket{\downarrow}$ spin state optically. Optimum cooling, assessed using Ramsey interferometry, occurs for a Raman drive at $\Omega=22\ \mathrm{MHz}$ and a resonant repump of $\Omega_{\mathrm{res}}=0.9 \Gamma_{0}/\sqrt{2}$ for an excited-state linewidth $\Gamma_{0}$, in agreement with the optimum conditions found in Ref. \cite{Gangloff2019}. These settings give an order-of-magnitude improvement in our electron spin inhomogeneous dephasing time $T_{2}^{*}$ (Fig. \ref{fig:2} (a)).
\subsubsection{Electron-spin control}
\begin{figure}
\includegraphics[width = 1\columnwidth,angle=0]{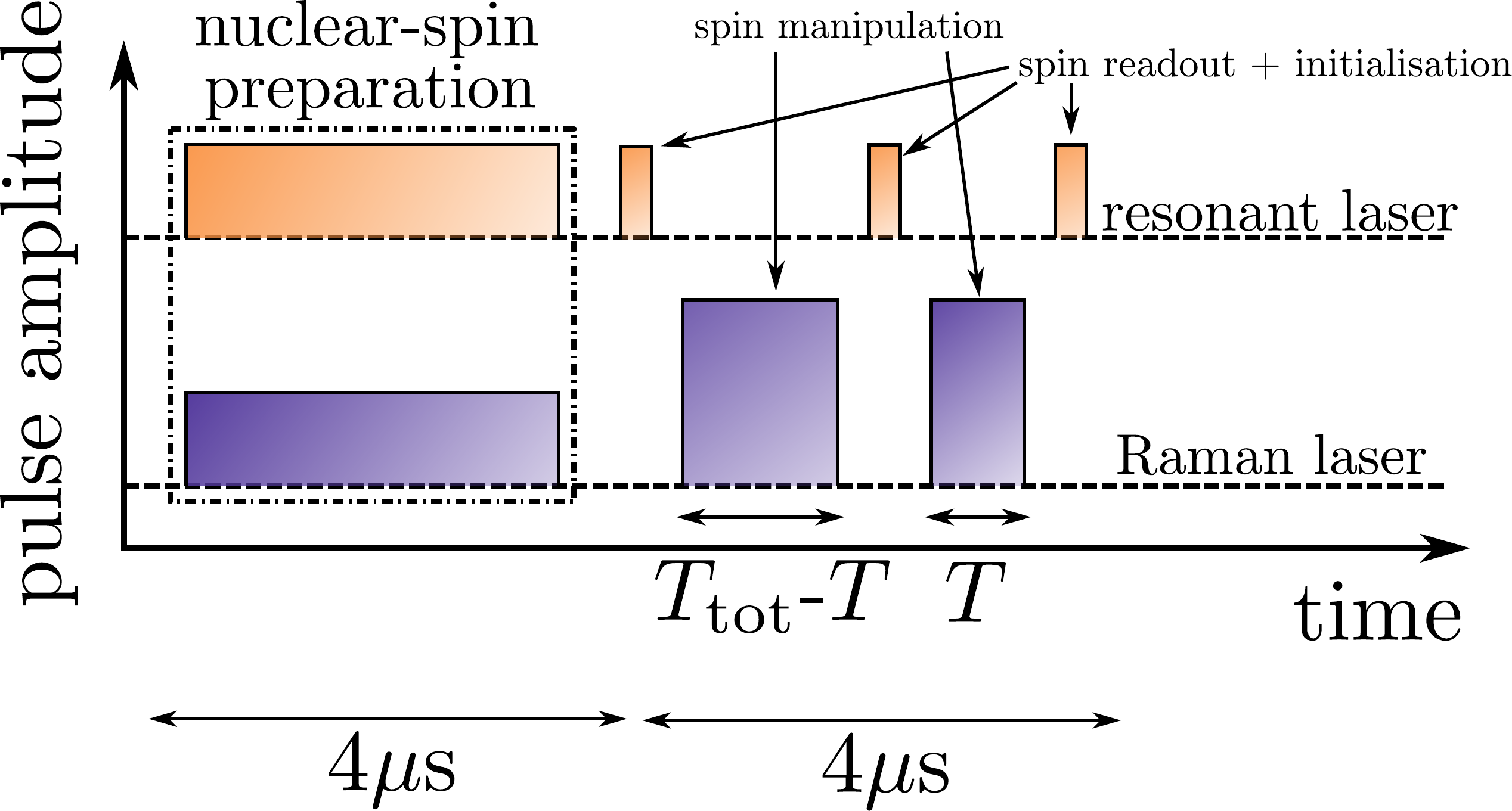}
\caption{A typical experimental cycle. The nuclear spin bath is prepared using a Raman cooling technique for 4$\mu$s, before electron spin control is performed. This takes place over a similar timescale, meaning that our duty cycle is close to 50\%.}
\label{fig:PulseSequence}
\end{figure}

During spin control, we conserve the total Raman pulse area in our sequences by pairing pulses of increasing length with pulses of decreasing length (Fig. \ref{fig:PulseSequence}). This allows us to stabilise the Raman laser power using a PID loop and maintain relative fluctuations below a per cent. We operate with a duty cycle of around 50\%, preparing the nuclear spin bath for a few $\mu$s before spending a similar amount of time performing electron spin control. The alternation on $\mathrm{\mu s}$ timescale of coherent manipulation and nuclear-spin preparation is fast compared with the nuclear-spin dynamics \cite{Ethier-Majcher2017} such that the nuclear-spin distribution is at steady state.
\section{Data Availability}
The data that support the plots within this paper and other findings of this study are
available from the corresponding author upon reasonable request.
\section{Acknowledgements}
This work was supported by the ERC PHOENICS grant (617985), the EPSRC Quantum Technology Hub NQIT (EP/M013243/1) and the Royal Society (RGF/EA/181068). D.A.G. acknowledges support from St John’s College Title A Fellowship.  E.V.D. acknowledges funding from the Danish Council for Independent Research (Grant No. DFF- 4181-00416). C.L.G. acknowledges support from a Royal Society Dorothy Hodgkin Fellowship.
\section{Author Contributions}
J.H.B., R.S., D.A.G., G.E.-M., D.M.J., C.L.G. and M.A. conceived the experiments. J.H.B., R.S. and C.L.G. acquired and analysed data. E.V.D., C.L.G. and J.H.B. developed the theory and performed simulations. E.C. and M.H. grew the sample. J.H.B., R.S., E.V.D., D.A.G., G.E.-M., D.M.J., C.L.G. and M.A. prepared the manuscript.

\clearpage
\setcounter{page}{1}
\setcounter{figure}{0}
\setcounter{equation}{0}
\renewcommand{\theequation}{S\arabic{equation}}
\renewcommand{\thefigure}{S\arabic{figure}}
\renewcommand{\bibnumfmt}[1]{[S#1]}
\renewcommand{\citenumfont}[1]{S#1}
\begin{center}
\textbf{\Large Supplemental Material}\\
\end{center}

\section{Further notes on the experimental setup}
\begin{figure}[h!]
\includegraphics[width=1\columnwidth,angle=0]{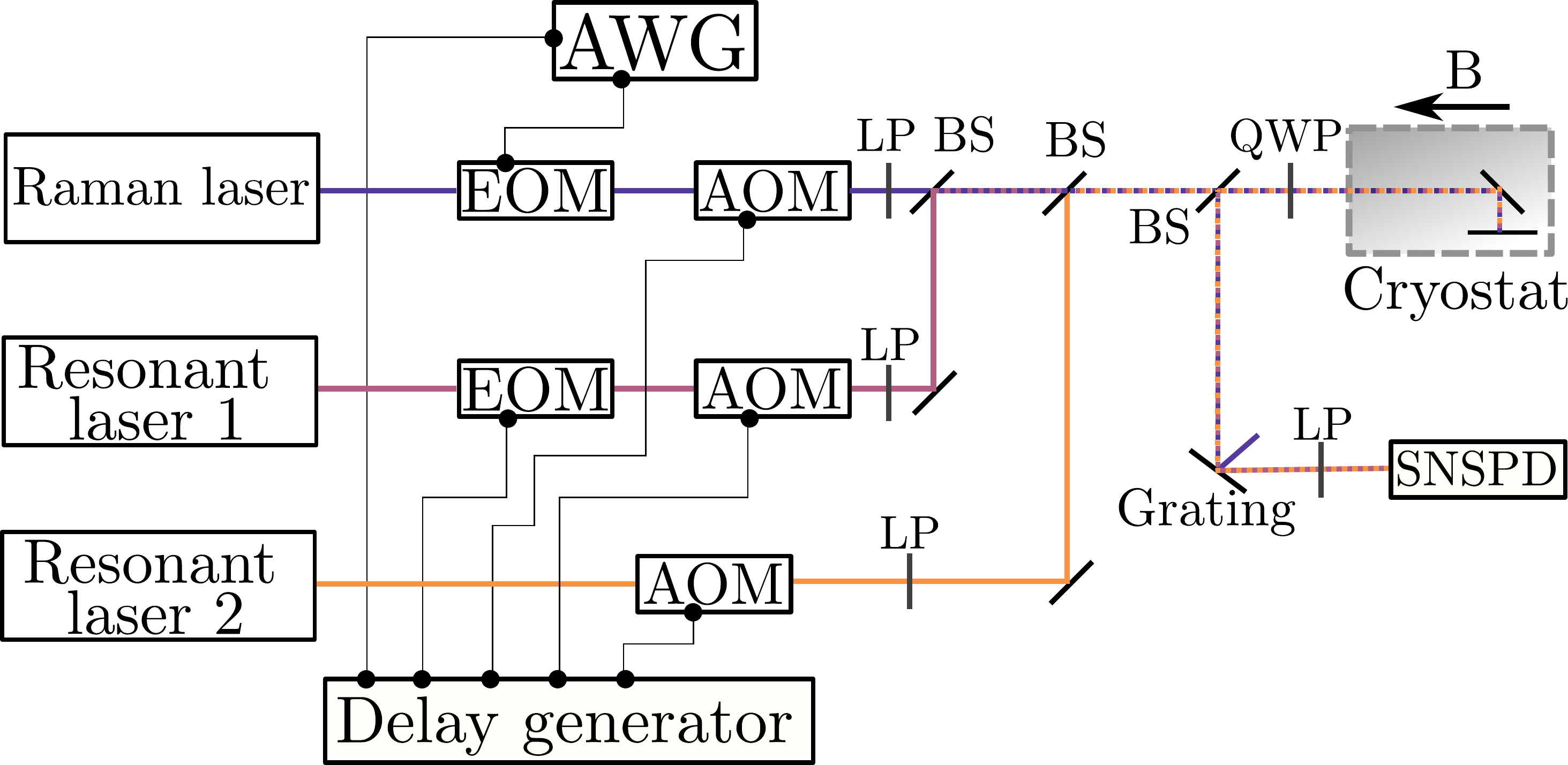}
\caption{A schematic of the experimental setup. Three CW lasers are passed through electro-optic modulators (EOMs) and acousto-optic modulators (AOMs) to build pulse sequences. The Raman-laser EOM is controlled by an arbitrary waveform generator (AWG). All other modulators are driven by delay generators, synchronised with the AWG. The lasers are combined using beamsplitters (BSs) and sent to a cryostat-housed QD device. Polarisation is controlled using a series of linear polarisers (LPs) and a quarter-wave plate (QWP), set such that the excitation is circularly polarised at the QD. Reflected laser-light is minimised using a cross-polarised detection. A grating further suppresses the Raman laser background. The filtered collection is sent to a superconducting nanowire single-photon detector (SNSPD).}
\label{fig:S1}
\end{figure}

Figure \ref{fig:S1} shows a schematic of the overall experimental setup. Three laser systems are combined and sent to the quantum dot (QD): a microwave-modulated Raman laser system (Toptica DL Pro, $\omega_R=2\pi\times 309300 \mathrm{\ GHz}$ ), a resonant laser to perform spin readout and initialisation (Newport NF laser, $\omega_1=2\pi\times 310051.2 \mathrm{\ GHz}$), and a second resonant laser to perform repump during the nuclear-spin cooling process (MogLabs CatEye laser, $\omega_1\approx \omega_1+\delta$, where $\delta$ compensates for a $~100-400 \mathrm{\ MHz}$ shift in the transition frequency induced by the Raman laser). The laser excitation and fluorescence collection is achieved using a confocal microscope with an $0.5$ NA objective lens. A cross-polarised detection minimises reflected resonant laser light, and a grating suppresses Raman laser light from the collection.\\
\section{Effective ESR frequency}
A QD in Voigt geometry has two excited states ($\ket{e}$ and $\ket{e'}$, split by the hole Zeeman energy $\omega_h$), giving rise to two paths for the Raman process \cite{SIPress2008}. These paths interfere and the polarisation of the Raman beams together with the phase-relationship between the optical transitions dictates the effective ESR Rabi frequency. The Raman laser is circularly polarised thus driving each arm of the $\Lambda$-levels with equal strength following $\Omega=\Omega_L^2/(2\Delta_L)$ where $\Omega_L$ is the optical Rabi frequency and $\Delta_L=\Delta\pm\omega_h/2$ ($\Delta$ is defined in Fig. 1 of the main text). The two Raman processes add up yielding an effective ESR frequency $\Omega=\Omega_L^2/\Delta$, in the limit $\omega_h\ll\Delta$.\\
\section{Rabi oscillations}
\subsection{Decay of Rabi oscillations limited by spin decay}
A resonantly driven 2-level system with a spin-decay process ($\Gamma_1$) which depolarises the electron spin can be described by the master equation:
\begin{equation}
\dot{\rho}=-i[\Omega\sigma_x,\rho]+\Gamma_1(L[\sigma_-]+L[\sigma_+])\rho,
\end{equation}
where $\sigma_x=\frac{1}{2}\ket{\uparrow}\bra{\downarrow}+\ket{\downarrow}\bra{\uparrow}$, $\sigma_+=\ket{\uparrow}\bra{\downarrow}$, $\sigma_-=\ket{\downarrow}\bra{\uparrow}$ and $L(a)\rho=a\rho a^{\dagger}-\frac{1}{2}\{a^{\dagger}a,\rho\}$.
The time evolution of the upper state population with the initial condition $\rho_{\uparrow\uparrow}(t=0)=1$ is:
\begin{equation}
\rho_{\uparrow\uparrow}(t)=\frac{1}{2}(1+e^{-3/2\Gamma_1 t}[\cos(\tilde{\Omega} t/2)-\frac{\Gamma_1}{\tilde{\Omega}}\sin(\tilde{\Omega} t/2)]),
\end{equation}
where $\tilde{\Omega}=\sqrt{4\Omega^2-\Gamma_1^2}$. Spin Rabi at ESR frequencies $\Omega>80MHz$ has a coherence limited by the extrinsic laser-induced spin-decay $\Gamma_1 \ll \Omega$, yielding a $1/e$-time $(3/2 \Gamma_1)^{-1}$ and a $Q$ factor $\approx \frac{4\Omega}{3\Gamma_1}$.\\
\subsection{Extraction of the $1/e$ time, $Q$ factor and pulse fidelity}
\begin{figure*}
\includegraphics[width=\textwidth,angle=0]{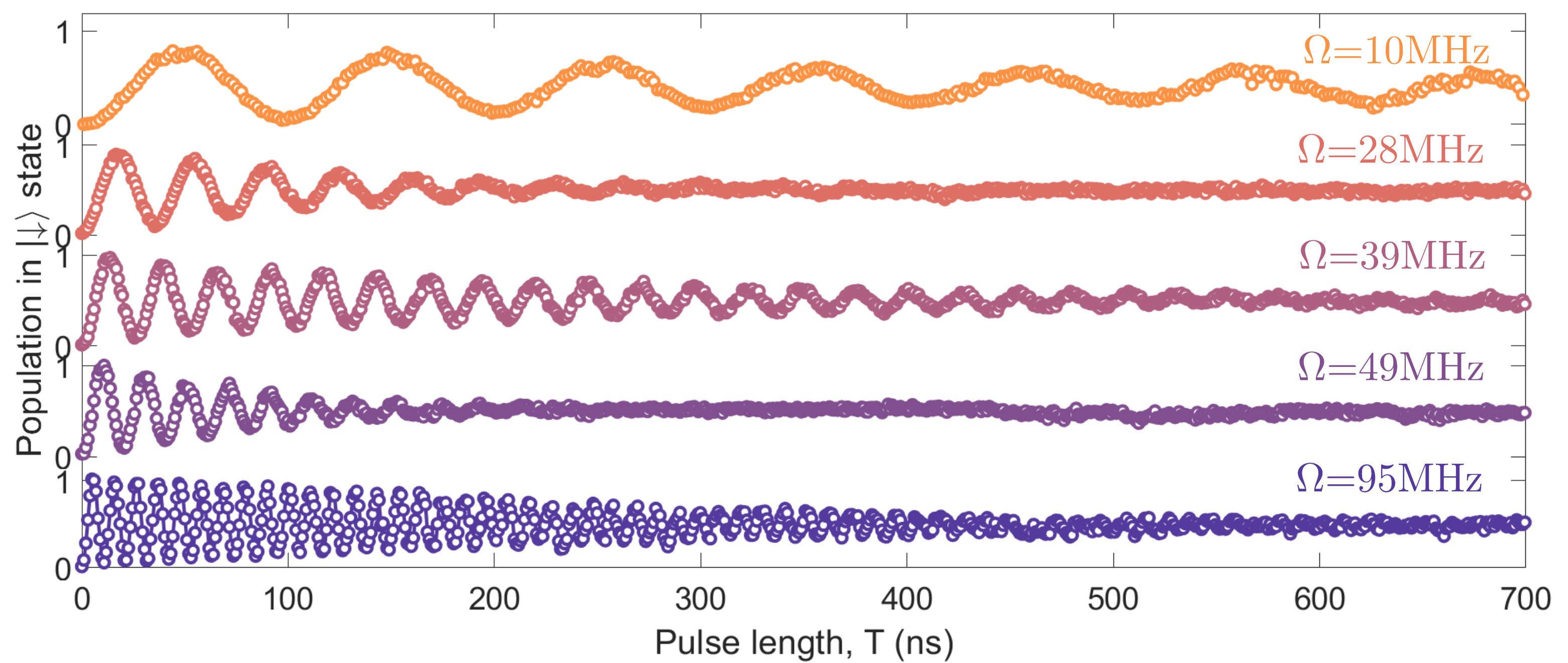}
\caption{Rabi oscillations at a set of different Rabi frequencies, illustrating the three regimes of (i) low Rabi frequency ($10\ \mathrm{MHz}$, top data), long decay time (ii) intermediate Rabi frequency ($28, 49\ \mathrm{MHz}$, 2nd and 4th data from top), very short decay time (iii) high Rabi frequency ($95\ \mathrm{MHz}$, bottom data), short decay time, as detailed in the main text. Data at $39\ \mathrm{MHz}$ (middle data) belongs to a region in the nuclear spectral density where coupling is low. These data were used to extract the decay times presented in the main text, where we measure up to a maximum pulse length of $790\ \mathrm{ns}$.}
\label{fig:LongRabi}
\end{figure*}
\begin{figure}
\includegraphics[width=1\columnwidth,angle=0]{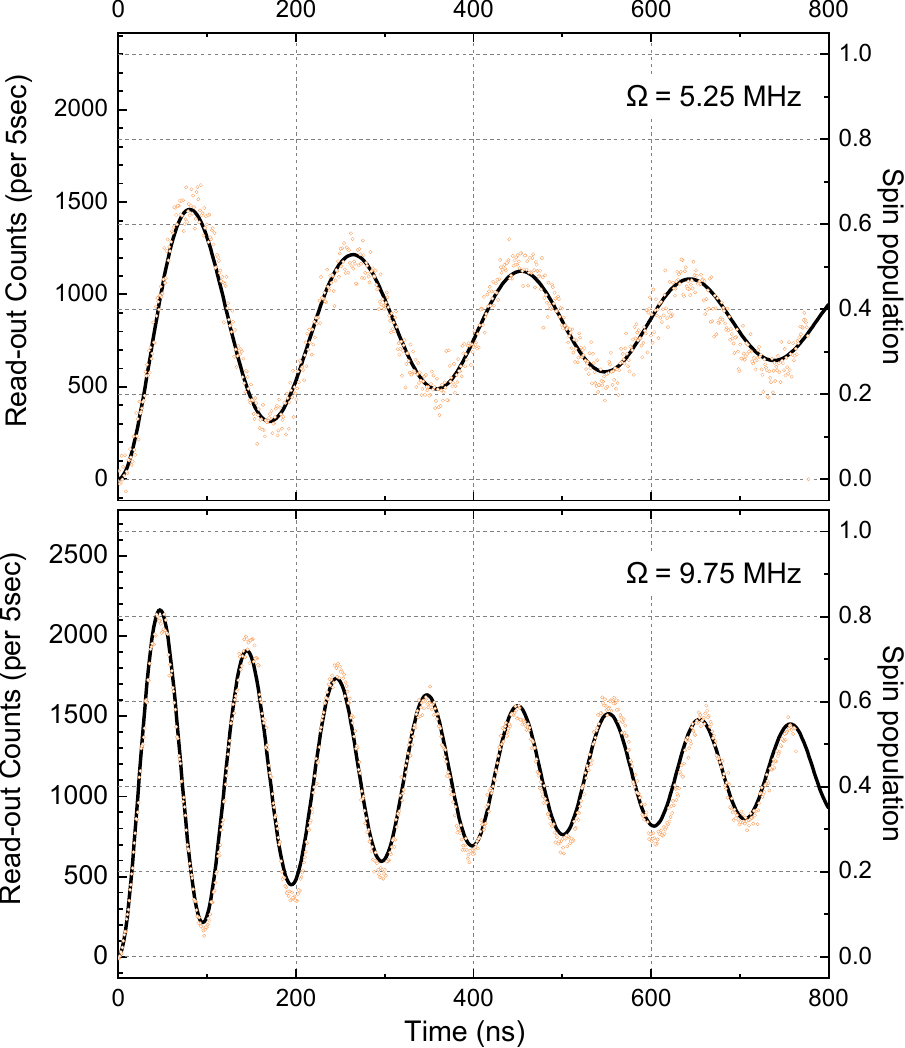}
\caption{Fidelity of ESR rotations at low Rabi frequencies. The spin population can be reconstructed from a two-level master equation model that accounts for nuclear-field inhomogeneities (black curve), yielding a $\pi$-pulse fidelity of $\sim 60\%$ at $\Omega\approx 5\mathrm{\ MHz}$ and $\sim 80\%$ at $\Omega\approx 10 \mathrm{\ MHz}$.}
\label{fig:LowPW}
\end{figure}

We measure Rabi oscillations as presented in Fig. \ref{fig:LongRabi}, up to ESR pulses of $790$ ns. For each dataset we evaluate the visibility over a $\pi$-period and measure the $1/e$ time, at which the visibility has decayed to $1/e$ of its initial value. The $Q$ factor is obtained as the ratio between this $1/e$ time and the $\pi$ pulse time [$t_{\pi}=1/(2\Omega)$].
In the high power regime ($\Omega\gg \omega^z_{\mathrm{nuc}}$), where the decay of the Rabi envelope is well-described by an exponential, the fidelity of a $\pi$ pulse is closely related to the $Q$ factor following $f_{\pi}=1/2(1+e^{-\frac{1}{Q}})$. In the low power regime  ($\Omega\ll \omega^z_{\mathrm{nuc}}$), the fidelity of a $\pi$ pulse can be obtained from fitting the Rabi oscillation to a two-level Bloch-equation model where we carry an averaging over a Gaussian detuning distribution of variance $\sigma_{\mathrm{OH}}=4.8\mathrm{\ MHz}$ (Fig. \ref{fig:LowPW}).\\
\subsection{Laser-induced spin decay}
\begin{figure}
\includegraphics[width=1\columnwidth,angle=0]{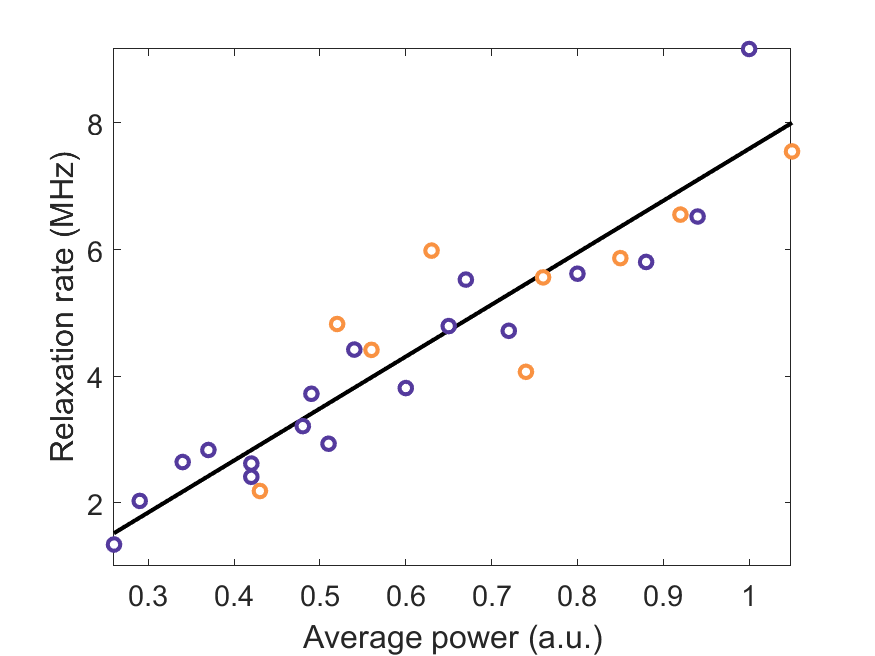}
\caption{Laser-induced spin relaxation as a function of input laser power, where the purple (yellow) circles indicate a laser detuning $\Delta$ of $800\ \mathrm{GHz}$ ($1600\ \mathrm{GHz}$). The black line is a linear fit to the data. The independence of relaxation rate from detuning indicates that this process is unrelated to the optical transitions of the QD.}
\label{fig:InducedRelaxation}
\end{figure}

At Rabi frequencies above $\sim 80\ \mathrm{MHz}$ (beyond the Hartmann-Hahn resonance), our decay envelope and corresponding gate fidelity become limited by laser-induced decay. At $\Delta=700\ \mathrm{GHz}$, the decay is $\sim 10^2$ times faster than the photon-scattering rate expected for ideal optical transitions (at our highest ESR drive $\Omega\sim 160 \mathrm{\ MHz}$, the photon-scattering rate is $2\Gamma_0 \Omega_L^2/\Delta_L^2 \approx 60$ kHz, where $\Gamma_0\sim140 \mathrm{\ MHz}$ is the optical linewidth). The identification of a laser-induced spin decay is further supported by pump-probe measurements presented in Fig. \ref{fig:InducedRelaxation}, where we measure the spin relaxation due to a detuned laser pulse (in the absence of any EOM modulation). The spin-relaxation rate increases linearly with the pulse power. If we increase the detuning (from $800\  \mathrm{GHz}$ to $1600\  \mathrm{GHz}$) but keep the power constant, we observe the same decay rate.\\
In previous work, it was proposed that incoherent processes such as trion dephasing led to the creation of excited state population. 
 However, the optical decoherence that has to be included to model the Rabi decay in Fig. 2 of the main text is incompatible with the close-to-lifetime-limited linewidth measured in resonance fluorescence. Phonons can also be ruled out both theoretically (we estimate phonon-absorption to be $10^2$-times smaller than off-resonant photon-scattering) and by our decay measurement (the exponentially suppressed phonon absorption beyond $k_BT\approx 80 \ \mathrm{GHz}$ would lead to very different decays at $800\ \mathrm{GHz}$ and $1600\ \mathrm{GHz}$ which is not the case for the decay observed here). Lastly, in our device, this laser-induced decay is even more pronounced for hole spins ($f_{\pi} \approx 0.92$ with ultrafast rotations or ESR rotations). Our observation of a detuning-independent laser-induced decay and qubit-dependent fidelities point towards non-resonant processes occurring directly within the ground-state manifold.\\
\section{Interactions with the nuclear-spin bath}
\label{sec:inter-with-nucl}
\subsection{Non-Markovian master equation}
\label{sec:non-markovian-master}
The Hamiltonian describing the driven central electron and the nuclear-spin bath, after a Schrieffer-Wolff transformation that yields the effective low-energy dynamics in the presence of lattice strain, is~\cite{SIDenning2019,SIGangloff2019},
\begin{align}
  \label{eq:1}
  H=H_\mathrm{e}+H_\mathrm{n}+H_\mathrm{hf}+H_\mathrm{nc},
\end{align}
where $H_\mathrm{e}$ describes the driven electron, $H_\mathrm{n}=\sum_j \omega_\mathrm{nuc}^zI_z^j + \Delta_Q^j (I_z^j)^2$ describes the free evolution of the nuclei, $H_\mathrm{hf}=\sum_{j}2A^jI_z^jS_z$ is the low-energy part of the hyperfine interaction and $H_\mathrm{nc}=-S_zV_\mathrm{n}$, with
\begin{align}
  \label{eq:2}
\begin{split}
  V_\mathrm{n}=&\sum_j\frac{A^jB_Q^j}{\omega_\mathrm{nuc}^z}\\
&\times \{[(I_x^j)^2-(I_y^j)^2]\cos^2\theta^j + [I_x^jI_z^j+I_z^jI_x^j]\sin2\theta^j\}
\end{split}
\end{align}
describes a non-collinear strain-induced hyperfine interaction. Here, $\omega_\mathrm{nuc}^z$ is the nuclear Zeeman splitting, $A^j, B_Q^j$ and $(\frac{\pi}{2}-\theta^j)$ are the hyperfine interaction strength, the quadrupolar coupling strength and the quadrupolar angle relative to the magnetic field for the $j$'th nucleus, and $\Delta_Q^j=B_Q^j(\sin^2\theta^j-\frac{1}{2}\cos^2\theta^j)$ is the associated quadrupolar energy shift.
The Overhauser field, $\Delta=\sum_j 2A^jI_z^j$ is modelled as a quasi-static classical variable~\cite{SIMerkulov2002} and is absorbed into $H_\mathrm{e}=\Omega S_x+\Delta S_z$. This non-interacting electron Hamiltonian can be diagonalised under the unitary transformation $H\rightarrow \tilde{H}=e^{i\phi S_y}He^{-i\phi S_y}$, where $\sin\phi=\Omega/\Omega',\; \cos\phi=\Delta/\Omega'$ and $\Omega'=\sqrt{\Omega^2+\Delta^2}$. The transformed terms in the Hamiltonian are then $\tilde{H}_\mathrm{e}=\Omega'S_z$, $\tilde{H}_\mathrm{n}=H_\mathrm{n}$, $\tilde{H}_\mathrm{nc}=(S_x\sin\phi-S_z\cos\phi)V_\mathrm{n}$.\\
 \indent To obtain the reduced dynamics of the electron spin density operator, $\rho$, we derive a quantum master equation, where the nuclear bath is traced out. When the system is operated in the vicinity of the Hartmann-Hahn resonance, the most significant contribution to the dynamics is expected to arise from the secular electron--nuclear transitions generated by $\tilde{H}_\mathrm{nc}$. Therefore, to simplify the analysis, we start out by removing the non-secular terms therein, obtaining $\tilde{H}\rightarrow \frac{1}{2}\sin\phi (S_- V_\mathrm{n}^+ + S_+V_\mathrm{n}^-)$, where
\begin{align}
  \label{eq:5}
  V_\mathrm{n}^+=\frac{1}{2}\sum_j A_\mathrm{nc}^j\qty[(I_+^j)^2\cos^2\theta_j+(I_+^jI_z^j+I_z^jI_+^j)\sin2\theta^j],
\end{align}
$A_\mathrm{nc}^j=\frac{A^jB_Q^j}{\omega_\mathrm{nuc}^z}$, $V_\mathrm{n}^-=(V_\mathrm{n}^+)^\dagger$ and $S_\pm=S_x\pm i S_y,\; I^j_\pm=I_x^j\pm iI_y^j$ are the electronic and nuclear spin transition operators.
The corresponding non-Markovian time-convolutionless master equation for $\rho$ is~\cite{SIBreuer2007}
\begin{align}
  \label{eq:6}
  \pdv{t}\rho=-i[\Omega'S_z,\rho] - \int_0^t\dd{\tau}\Tr_\mathrm{n}[\tilde{H}_\mathrm{nc},[\tilde{H}_\mathrm{nc}(-\tau),\rho\otimes \rho_\mathrm{n}^0]]
\end{align}
where $\tilde{H}_\mathrm{nc}(-\tau)=e^{-i(\tilde{H}_\mathrm{e}+\tilde{H}_\mathrm{n})\tau}\tilde{H}_\mathrm{nc}e^{+i(\tilde{H}_\mathrm{e}+\tilde{H}_\mathrm{n})\tau}$ denotes the interaction picture time evolution of $\tilde{H}_\mathrm{nc}$ and $\rho_\mathrm{n}^0$ is the reference state of the nuclear bath. Following Ref.~\cite{SICoish2010}, we assume that the nuclear reference state is factorisable among the nuclei. Furthermore, we assume that the relevant features contributing to the non-collinear processes in the master equation, Eq.~\eqref{eq:6}, can be described by a thermal nuclear density operator at infinite temperature. Under these assumptions, we arrive at the following master equation for the electron spin,
\begin{align}
  \label{eq:7}
  \pdv{t}\rho(t)=-i[\Omega'S_z,\rho(t)] + \Gamma(\Omega',t)(L(S_+) + L(S_-) )\rho(t),
\end{align}
where the nuclear-induced Lamb shift has been neglected, $L(x)\rho=x^\dagger \rho x - \frac{1}{2}\{x x^\dagger, \rho \}$ is the Lindblad dissipator and
\begin{align}
  \label{eq:8}
\Gamma(\Omega',t)= \frac{\sin^2\phi}{4}\frac{1}{\pi}\int \dd{\omega} \mathcal{D}(\omega)\frac{\sin[(\omega-\Omega')t]}{\omega-\Omega'}
\end{align}
is a time-dependent decay rate calculated from the spectral density, $\mathcal{D}(\omega)=\mathcal{D}^{(1)}(\omega)+\mathcal{D}^{(2)}(\omega)$, which contains contributions from the nuclear processes changing the total nuclear polarisation by one or two units,
\begin{align}
  \label{eq:9}
\begin{split}
  &\mathcal{D}_1(\omega)=\frac{\pi}{2}\sum_j \frac{(A_\mathrm{nc}^j \sin 2\theta^j)^2}{2I^j+1}\\
& \sum_{m_j=-I^j}^{I^j-1} [M_+(I^j,m_j)(2m_j+1)]^2 \\
&\times\delta(\omega-[\omega_\mathrm{nuc}^z+(2m_j+1)\Delta_Q^j]), \\
&\ \\
&\mathcal{D}_2(\omega)=\frac{\pi}{2}\sum_j \frac{(A_\mathrm{nc}^j\cos^2\theta^j)^2}{2I^j+1} \\
&\sum_{m_j=-I^j}^{I^j-2} [M_+(I^j,m_j)M_+(I^j,m_j+1)]^2 \\
&\times \delta(\omega-[2\omega_\mathrm{nuc}^z+4\Delta_Q^j(m_j+1)]),
\end{split}
\end{align}
where $M_+(I,m)=\sqrt{I(I+1)-m(m+1)}$ and $I^j$ is the total spin eigenvalue for the $j$'th nucleus. The next step is to split the summation over nuclei into a summation over nuclear species, $s$, such that $\mathcal{D}^{(i)}=\sum_s \mathcal{D}^{(i)}_s$. For each species, the total nuclear spin is constant, $I^j=I_s$, and the parameters $(\theta,B_Q,A)=:\xi$ are described by a statistical distribution over the nuclear ensemble, $P_s(\xi)$, for the given species, $s$. We then approximate the summation over nuclei in Eq.~\eqref{eq:9} as an integral over this distribution, $\sum_j f_s^j\simeq N_s\int\dd{\xi}P_s(\xi)f_s(\xi)$, where $N_s$ is the number of nuclei of species $s$ and $f^j_s$ is a general function of single-nucleus parameters of that species. Taking the distribution $P(\xi)$ to be factorisable, $P_s(\xi)=p_{s,1}(\theta)p_{s,2}(B_Q)p_{s,3}(A)$, we find
\begin{align}
  \label{eq:10}
\begin{split}
  \mathcal{D}_s^{(1)}(\omega)&=\frac{\pi}{2}\frac{\langle A^2\rangle_s N_s}{2I_s+1}\sum_{m=-I_s}^{I_s-1}[M_+(I_s,m)(2m+1)]^2 \\
&\times \int\dd{\theta} p_1(\theta)p_2\qty[\frac{\omega-\omega_\mathrm{nuc}^z}{(2m+1)(\sin^2\theta-\frac{1}{2}\cos^2\theta)}] \\
&\times \qty(\frac{(\omega-\omega_\mathrm{nuc}^z)\sin2\theta}{\omega_\mathrm{nuc}^z(2m+1)(\sin^2\theta-\frac{1}{2}\cos^2\theta)})^2 \\
&\times \abs{(2m+1)\qty(\sin^2\theta-\frac{1}{2}\cos^2\theta)}^{-1} \\
\end{split}
\end{align}

\begin{align*}
\begin{split}
\mathcal{D}_s^{(2)}(\omega)&=\frac{\pi}{4}\frac{\langle A^2\rangle_s N_s}{2I_s+1}\sum_{m=-I_s}^{I_s-2}[M_+(I_s,m)M_+(I_s,m+1)]^2 \\
&\times\int\dd{\theta}p_1(\theta) p_2\qty[\frac{\omega-2\omega_\mathrm{nuc}^z}{4(m+1)(\sin^2\theta-\frac{1}{2}\cos^2\theta)}] \\
&\times \qty(\frac{(\omega-2\omega_\mathrm{nuc}^z)\cos^2\theta}{4\omega_\mathrm{nuc}^z(m+1)(\sin^2\theta-\frac{1}{2}\cos^2\theta)})^2\\
&\times \abs{2(m+1)\qty(\sin^2\theta-\frac{1}{2}\cos^2\theta)}^{-1},
\end{split}
\end{align*}
where $\ev{A^2}_s=\int\dd{A}p_{s,3}(A)A^2$.

Transforming back to the Zeeman eigenbasis, the master equation is 
\begin{align}
  \label{eq:14}
\begin{split}
  \pdv{t}\rho(t)=&-i[\Delta S_z+\Omega S_x,\rho(t)] \\ &+ \Gamma(\Omega',t)\{L(S_\phi)+L(S_\phi^\dagger)\}\rho(t),
\end{split}
\end{align}
where $S_\phi=S_x\cos\phi +i S_y+  S_z\sin\phi$.
Finally, we add the terms $\Gamma_1(L(S_+)+L(S_-))\rho(t)$ and $\Gamma_2 L(S_z)$ to the master equation, where $\Gamma_1$ is the extrinsic laser-induced spin-decay process and $\Gamma_2$ is the spin coherence decay measured in Hahn-Echo, $1/(2.8\mathrm{\ \mu s})$.

\subsection{Parameter probability distributions}
\label{sec:param-prob-distr}

The probability distributions for the hyperfine and quadrupolar coupling strengths, $p_{s,2}$ and $p_{s,3}$ are taken Gaussian. The major quadrupolar axis distribution is assumed to be symmetric around the QD growth axis, characterised by a uniform distribution of the azimuthal angle, $\varphi'$ and a Gaussian distribution for the polar angle, $\theta'$. The equivalent distribution for the $\theta$-angle appearing in Eq.~\eqref{eq:2} is obtained by rotating the coordinate system around the magnetic field axis (the $x$-axis), such that the quadrupolar angle is lying in the $xz$-plane. Denoting the Gaussian polar probability distribution for the nuclear species $s$ by $p_{s,\mathrm{p}}(\theta')$, the distribution for $\theta$ is found to be
\begin{align}
  \label{eq:11}
  p_{s,1}(\theta)=\frac{1}{\pi}\int_{\theta}^{\pi-\theta}\dd{\theta'} \frac{p_{s,\mathrm{p}}(\theta')\cos(\theta)\sin\theta'}{\sqrt{\sin^2\theta'-\sin^2\theta}},
\end{align}
where $\theta$ is defined to be in the range $[0,\pi]$.

\subsection{Rabi decay rate}
\label{sec:rabi-decay-rate}

Due to the non-Markovianity of the electron spin time evolution, a decay rate of the Rabi oscillations is in principle not well-defined. However, the non-Markovian effects are most strongly pronounced at short times, whereas in the long-time limit, the system approaches the Markovian limit. Effectively, the electron spin probes the spectral density at the Rabi frequency during a finite time window corresponding to the decay time. This can be encoded into the calculation of the dynamics by employing a self-consistent Born-Markov approximation~\cite{SIEsposito2010,SIJin2014}. Here, we implement such an approach by first writing the Markov limit for the nuclear transition induced electron decay rate,
\begin{align}
  \label{eq:3}
  \Gamma_\mathrm{M}(\Omega')=\frac{1}{4}\sin^2\phi\Re\qty[\int_0^\infty\dd{\tau}e^{-i\Omega'\tau}\int_{-\infty}^\infty \frac{\dd{\omega}}{2\pi}\mathcal{D}(\omega)e^{i\omega\tau}].
\end{align}
Here, the exponential factor $e^{-i\Omega'\tau}$ appears through the free evolution of the electronic $S_\pm$ operators. In our self-consistent Born-Markov approach, we encode the decay of the electron spin into this correlation function, replacing it by $e^{-[i\Omega'+\gamma(\Omega')]t}$. The damping rate, $\gamma(\Omega')$, is then determined self-consistently through an iterative process. By replacing the free correlation function by the damped one, we define a self-consistent Markovian decay rate,
\begin{align}
  \label{eq:4}
\begin{split}
  \Gamma_\mathrm{SCM}(\Omega')&= \frac{\sin^2 \phi}{4}\times 2\Re\Big[\int_0^\infty\dd{\tau}e^{-[i\Omega'+\gamma(\Omega')]\tau} \\ &\hspace{0.2\columnwidth}\times\int_{-\infty}^\infty\frac{\dd{\omega}}{2\pi}\mathcal{D}(\omega)e^{i\omega\tau}\Big] \\
&= \frac{\sin^2 \phi}{4} \int_{-\infty}^\infty\dd{\omega} \mathcal{D}(\omega) \frac{1}{\pi}\frac{\gamma(\Omega')}{\gamma(\Omega')^2 + (\omega-\Omega')^2},
\end{split}
\end{align}
which describes a convolution of the spectral density with a Lorentzian distribution. Furthermore, we also average over the configurations of the Overhauser field, which is taken as a Gaussian distribution with standard deviation $\sigma_\mathrm{OH}$, leading to the averaged decay rate
\begin{align}
  \label{eq:13}
\begin{split}
  \tilde{\Gamma}_\mathrm{SCM}(\Omega)&=\frac{1}{4}\int\dd{\Delta}\int_{-\infty}^\infty\dd{\omega} \frac{\Omega^2}{\Delta^2+\Omega^2} \mathcal{D}(\omega)
\\ &\times \frac{e^{-\Delta^2/2\sigma_\mathrm{OH}^2}}{\sqrt{2\pi\sigma_\mathrm{OH}^2}}\frac{1}{\pi}\frac{\gamma(\Omega')}{\gamma(\Omega')^2 + (\omega-\Omega')^2}.
\end{split}
\end{align}

To self-consistently determine $\tilde{\Gamma}_\mathrm{SCM}(\Omega)$ and $\gamma(\Omega')$, we start out by letting $\gamma(\Omega')=\frac{1}{4}\mathcal{D}(\Omega)+\frac{3}{2}\Gamma_1+\Gamma_2$ and calculate the first iteration of the decay rate, $\tilde{\Gamma}^{(1)}_\mathrm{SCM}(\Omega)$. In the next iteration, we set $\gamma(\Omega')=\tilde{\Gamma}^{(1)}_\mathrm{SCM}(\Omega)+\frac{3}{2}\Gamma_1+\Gamma_2$ and repeat this procedure until the iterative series has converged.  


\end{document}